\documentclass{article}
\usepackage{amssymb}
\usepackage{amsfonts}
\usepackage{amsmath}

\newtheorem{theorem}{Theorem}
\newtheorem{acknowledgement}[theorem]{Acknowledgement}

\input{tcilatex}
\begin{document}

\title{Brane gravity in $4D$ from Chern-Simons gravity theory}
\author{R. D\'{\i}az$^{1,}$\thanks{%
roberto.diaz@ulagos.cl } , F. G\'{o}mez$^{1,}$\thanks{%
fernando.gomez@ulagos.cl} , M. Pinilla$^{1,}$\thanks{%
merardo.pinilla@ulagos.cl } \ and P. Salgado$^{2}$\thanks{%
patsalgado@unap.cl} \\
$^{1}$Departamento de Ciencias Exactas, Universidad de Los Lagos, \\
Avenida Fuchslocher 1305, Osorno, Chile\\
$^{2}$Facultad de Ciencias, Universidad Arturo Prat\\
Avenida Arturo Prat 2120, Iquique, Chile }
\maketitle

\begin{abstract}
We evaluate a $5$-dimensional Randall Sundrum type metric in the Lagrangian
of the Einstein-Chern-Simons gravity, and then we derive an action and its
corresponding field equations, for a $4$-dimensional brane embedded in the $%
5 $-dimensional space-time of the theory, which in the limit $l\rightarrow 0$
leads to the $4$-dimensional general relativity with cosmological constant.
An interpretation of the $h^{a}$ matter field present in the
Einstein-Chern-Simons gravity action is given. As an application, we find
some Friedmann-Lemaitre-Robertson-Walker cosmological solutions that exhibit
accelerated behavior.
\end{abstract}

\section{\textbf{Introduction}}

The study of the so called brane world models has introduced completely new
ways of looking upon standard problems in many areas of theoretical physics.
The existence of those new dimensions may have non trivial effects in our
understanding of the cosmology of the early Universe, among many other
issues.

The idea dates back to the $1920\prime $s, to the works of Kaluza and Klein 
\cite{kaluza,klein} who tried to unify electromagnetism with Einstein
gravity by assuming that the photon originates from the fifth component of
the metric.\ By convention, it has always been assumed that such extra
dimensions should be compactified to manifolds of small radii with sizes of
the order of the Planck length.

It was only in the last years of the $20th$ century when people started to
ask the question of how large could these extra dimensions be without
getting into conflict with observations.

Of particular interest in this context are the Randall and Sundrum models 
\cite{randall,randall1} for warped backgrounds, with compact or even
infinite extradimensions. Randall and Sundrum proposed that the metric of
the space-time is given by

\begin{equation}
ds^{2}=e^{-2kr_{c}}\eta _{\mu \nu }dx^{\mu }dx^{\nu }+r_{c}^{2}d\phi ^{2},
\label{1}
\end{equation}%
i.e. a $4$-dimensional metric multiplied by a "warp\ factor" which is a
rapidly changing function of an additional dimension, where $k$ is a scale
of order the Planck scale, $x^{\mu }$ are coordinates for the familiar four
dimensions, while $0\leq \phi \leq \pi $ is the coordinate for an extra
dimension, which is a finite interval whose size is set by $r_{c}$, known as
"compactification radius".\ Randall and Sundrum showed that this metric is a
solution to Einstein's equations.

The Einstein-Chern-Simons gravity \cite{salg1} is a gauge theory whose
Lagrangian density is given by a 5-dimensional Chern-Simons form for the so
called $\mathfrak{B}$ algebra. This algebra can be obtained from the Anti-de
Sitter algebra and a particular semigroup $S$ by means of the $S$-expansion
procedure introduced in Refs. \cite{salg2,salg3}.\ The field content induced
by the $\mathfrak{B}$ algebra includes the vielbein $e^{a}$, the spin
connection $\omega ^{ab}$, and two extra bosonic fields $h^{a}$ and $k^{ab}.$
The Einstein-Chern-Simons gravity has the interesting property that the
5-dimensional Chern-Simons Lagrangian for the $\mathcal{B}$ algebra, given
by \cite{salg1}%
\begin{eqnarray}
L_{ChS}^{(5)}[e,\omega ,h,k] &=&\alpha _{1}l^{2}\varepsilon
_{abcde}R^{ab}R^{cd}e^{e}  \label{1t} \\
&&+\alpha _{3}\varepsilon _{abcde}\left( \frac{2}{3}%
R^{ab}e^{c}e^{d}e^{e}+2l^{2}k^{ab}R^{cd}T^{\text{ }e}+l^{2}R^{ab}R^{cd}h^{e}%
\right) ,  \notag
\end{eqnarray}%
where $R^{ab}=\mathrm{d}\omega ^{ab}+\omega _{\text{ }c}^{a}\omega ^{cb}$
and $T^{a}=\mathrm{d}e^{a}+\omega _{\text{ }c}^{a}e^{c}$, leads to the
standard general relativity without cosmological constant in the limit where
the coupling constant $l$ tends to zero while keeping the Newton's constant
fixed. \ It should be noted that there is an absence of kinetic terms for
the fields $h^{a}$ and $k^{ab}$ in the Lagrangian $L_{ChS}^{(5)}$ \cite%
{salg4} (for detail see appendix).

The main purpose of this letter is to make the $5$-dimensional
Einstein-Chern-Simons theory consistent with the idea of a $4$-dimensional
space-time, through the replacement of a Randall Sundrum type metric in the
Lagrangian (\ref{1t}).\ It is also a purpose of this paper to get an
interpretation of the $h^{a}$ matter field present in the
Einstein-Chern-Simons gravity action.

The article is organized as follows: in section 2, we evaluate a $5$%
-dimensional Randall Sundrum type metric in the Lagrangian (\ref{1t}), and
then we derive an action for a $4$-dimensional brane embedded in the $5$%
-dimensional space-time of the Einstein-Chern-Simons gravity, which in the
limit $l\rightarrow 0$\ leads to the $4$-dimensional general relativity with
cosmological constant. In section $3$, we write the action for the brane in
tensorial language and then we find the corresponding Friedmann equations
for homogeneous and isotropic cosmology, and we obtain some De Sitter
accelerated solutions. Finally, concluding remarks are presented in section $%
4$.

\section{\textbf{Action for a }$4$\textbf{-dimensional brane embedded in the 
}$5$\textbf{-dimensional space-time}}

Chern--Simons theories of gravity are valid only in odd-dimensions and, in
order to have a well defined even-dimensional theory, it would be necessary
to carry out some kind of dimensional reduction or compactification.

In Refs. \cite{randall,randall1} a mechanism was introduced that allows to
connect the Einstein-Chern--Simons action with an action for a $4$%
-dimensional brane embedded in the $5$-dimensional space-time. \ 

In Ref. \cite{salg1}, the $5$-dimensional Chern-Simons Lagrangian for
gravity (\ref{1t}) was developed. The corresponding action is invariant
under the so-called $\mathfrak{B}$ algebra, which induces two geometrical
fields, the vielbein $e^{a}$ and the spin connection $\omega ^{ab}$, and two
bosonic matter fields, $h^{a}$ and $k^{ab}$. In order to find an action and
its corresponding field equations for a $4$-dimensional brane embedded in
the $5$-dimensional space-time $\Sigma _{5}$ of the Einstein-Chern-Simons
gravity, we will consider the following $5$-dimensional Randall Sundrum type
metric%
\begin{eqnarray}
ds^{2} &=&e^{2f(\phi )}\tilde{g}_{\mu \nu }(\tilde{x})d\tilde{x}^{\mu }d%
\tilde{x}^{\nu }+r_{c}^{2}d\phi ^{2}  \notag \\
&=&e^{2f(\phi )}\tilde{\eta}_{mn}\tilde{e}^{m}\tilde{e}^{n}+r_{c}^{2}d\phi
^{2},  \label{5t}
\end{eqnarray}%
where $e^{2f(\phi )}$ is the so-called "warp factor", and $r_{c}$ is the
so-called "compactification radius" of the extra dimension, which is
associated with the coordinate $0\leqslant \phi <2\pi $. The symbol $\sim $
denotes $4$-dimensional quantities related to the space-time $\Sigma _{4}.$
We will use the usual notation, 
\begin{eqnarray}
x^{\alpha } &=&\left( \tilde{x}^{\mu },\phi \right) ;\text{ \ \ \ \ }\alpha
,\beta =0,...,4;\text{ \ \ \ \ }a,b=0,...,4;  \notag \\
\mu ,\nu &=&0,...,3;\text{ \ \ \ \ }m,n=0,...,3;  \notag \\
\eta _{ab} &=&diag(-1,1,1,1,1);\text{ \ \ \ \ }\tilde{\eta}%
_{mn}=diag(-1,1,1,1).  \label{6t}
\end{eqnarray}

This allows us, for example, to write 
\begin{equation}
e^{m}(\phi ,\tilde{x})=e^{f(\phi )}\tilde{e}^{m}(\tilde{x})=e^{f(\phi )}%
\tilde{e}_{\text{ }\mu }^{m}(\tilde{x})d\tilde{x}^{\mu };\text{ \ \ }%
e^{4}(\phi )=r_{c}d\phi .\text{\ }  \label{s2}
\end{equation}

From the vanishing torsion condition%
\begin{equation}
T^{a}=de^{a}+\omega _{\text{ }b}^{a}e^{b}=0,  \label{2t}
\end{equation}%
we obtain 
\begin{equation}
\omega _{\text{ }b\alpha }^{a}=-e_{\text{ }b}^{\beta }\left( \partial
_{\alpha }e_{\text{ }\beta }^{a}-\Gamma _{\text{ }\alpha \beta }^{\gamma }e_{%
\text{ }\gamma }^{a}\right) ,  \label{3t}
\end{equation}%
where $\Gamma _{\text{ }\alpha \beta }^{\gamma }$ is the Christoffel symbol.

From equations (\ref{s2}) and (\ref{2t}), we find%
\begin{equation}
\omega _{\text{ }4}^{m}=\frac{e^{f}f^{\prime }}{r_{c}}\tilde{e}^{m},\text{
with }f^{\prime }=\frac{\partial f}{\partial \phi },  \label{102t}
\end{equation}%
and the $4$-dimensional vanishing torsion condition 
\begin{equation}
\tilde{T}^{m}=\tilde{d}\tilde{e}^{m}+\tilde{\omega}_{\text{ }n}^{m}\tilde{e}%
^{n}=0,\text{ with \ }\tilde{\omega}_{\text{ }n}^{m}=\omega _{\text{ }n}^{m}%
\text{ \ and }\tilde{d}=d\tilde{x}^{\mu }\frac{\partial }{\partial \tilde{x}%
^{\mu }}.  \label{1030t}
\end{equation}

From (\ref{102t}), (\ref{1030t}) and the Cartan's second structural
equation, $R^{ab}=d\omega ^{ab}+\omega _{\text{ }c}^{a}\omega ^{cb}$, we
obtain the components of the $2$-form curvature%
\begin{equation}
R^{m4}=\frac{e^{f}}{r_{c}}\left( f^{\prime 2}-f^{\prime \prime }\right)
d\phi \tilde{e}^{m},\text{ \ }R^{mn}=\tilde{R}^{mn}-\left( \frac{%
e^{f}f^{\prime }}{r_{c}}\right) ^{2}\tilde{e}^{m}\tilde{e}^{n},\text{\ }
\label{105t}
\end{equation}%
where the $4$-dimensional $2$-form curvature is given by%
\begin{equation}
\tilde{R}^{mn}=\tilde{d}\tilde{\omega}^{mn}+\tilde{\omega}_{\text{ }p}^{m}%
\tilde{\omega}^{pn}.
\end{equation}

The torsion-free condition implies that the third term in the equation (\ref%
{1t}) vanishes. This means that the Lagrangian (\ref{1t}) is no longer
dependent on the field $k^{ab}$. So that the Lagrangian (\ref{1t}) has two
independent fields, $e^{a}$ and $h^{a}$, and it is given by 
\begin{equation}
L_{ChS}^{(5)}[e,h]=\alpha _{1}l^{2}\varepsilon
_{abcde}R^{ab}R^{cd}e^{e}+\alpha _{3}\varepsilon _{abcde}\left( \frac{2}{3}%
R^{ab}e^{c}e^{d}e^{e}+l^{2}R^{ab}R^{cd}h^{e}\right) .  \label{4t}
\end{equation}

From equation (\ref{4t}) we can see that the Lagrangian contains the
Gauss-Bonnet term $L_{GB}$, the Einstein-Hilbert term $L_{EH}$ and a term $%
L_{H}$ which couples geometry and matter. In fact, replacing (\ref{s2}) and (%
\ref{105t}) in (\ref{4t}), and using $\tilde{\varepsilon}_{mnpq}=\varepsilon
_{mnpq4}$, we obtain%
\begin{eqnarray}
L_{GB} &=&\varepsilon _{abcde}R^{ab}R^{cd}e^{e}  \notag \\
&=&r_{c}d\phi \left \{ \tilde{\varepsilon}_{mnpq}\tilde{R}^{mn}\tilde{R}%
^{pq}-\left( \frac{2e^{2f}}{r_{c}^{2}}\right) \left( 3f^{\prime
2}+2f^{\prime \prime }\right) \tilde{\varepsilon}_{mnpq}\tilde{R}^{mn}\tilde{%
e}^{p}\tilde{e}^{q}\right.  \notag \\
&&\text{ \ \ \ \ \ \ \ \ \ \ \ \ \ \ }\left. +\left( \frac{e^{4f}}{r_{c}^{4}}%
f^{\prime 2}\right) \left( 5f^{\prime 2}+4f^{\prime \prime }\right) \tilde{%
\varepsilon}_{mnpq}\tilde{e}^{m}\tilde{e}^{n}\tilde{e}^{p}\tilde{e}%
^{q}\right \} ,  \label{8t}
\end{eqnarray}

\begin{eqnarray}
L_{EH} &=&\varepsilon _{abcde}R^{ab}e^{c}e^{d}e^{e}  \notag \\
&=&r_{c}d\phi \left\{ \left( 3e^{2f}\right) \tilde{\varepsilon}_{mnpq}\tilde{%
R}^{mn}\tilde{e}^{p}\tilde{e}^{q}\right.  \notag \\
&&-\left. \left( \frac{e^{4f}}{r_{c}^{2}}\right) \left( 5f^{\prime
2}+2f^{\prime \prime }\right) \tilde{\varepsilon}_{mnpq}\tilde{e}^{m}\tilde{e%
}^{n}\tilde{e}^{p}\tilde{e}^{q}\right\} ,  \label{9t}
\end{eqnarray}

\begin{eqnarray}
L_{H} &=&\varepsilon _{abcde}R^{ab}R^{cd}h^{e}  \notag \\
&=&\tilde{\varepsilon}_{mnpq}\left \{ \tilde{R}^{mn}-\left( \frac{%
e^{f}f^{\prime }}{r_{c}}\right) ^{2}\tilde{e}^{m}\tilde{e}^{n}\right \}
\label{10t} \\
&&\text{ \ \ \ \ \ \ }\times \left \{ \left[ \tilde{R}^{pq}-\left( \frac{%
e^{f}f^{\prime }}{r_{c}}\right) ^{2}\tilde{e}^{p}\tilde{e}^{q}\right]
h^{4}-\left( \frac{4e^{f}}{r_{c}}\right) \left( f^{\prime 2}+f^{\prime
\prime }\right) d\phi \tilde{e}^{p}h^{q}\right \} .  \notag
\end{eqnarray}

Note that in equation (\ref{8t}) there is a quadratic term in the curvature
given by $r_{c}d\phi \tilde{\varepsilon}_{mnpq}\tilde{R}^{mn}\tilde{R}^{pq}$%
. The action associated with this term can be directly integrated\ in $\phi $%
.\ In fact,%
\begin{equation}
r_{c}\int_{\Sigma _{5}}d\phi \tilde{\varepsilon}_{mnpq}\tilde{R}^{mn}\tilde{R%
}^{pq}=r_{c}\int_{0}^{2\pi }d\phi \int_{\Sigma _{4}}\tilde{\varepsilon}%
_{mnpq}\tilde{R}^{mn}\tilde{R}^{pq}=2\pi r_{c}\int_{\Sigma _{4}}\tilde{%
\varepsilon}_{mnpq}\tilde{R}^{mn}\tilde{R}^{pq},
\end{equation}%
which represents the $4$-dimensional Gauss-Bonnet term. This term is a
topological term, so that it does not contribute to the dynamics, and it can
be eliminated.

There is also a quadratic term in the curvature in equation (\ref{10t}): $%
\tilde{\varepsilon}_{mnpq}\tilde{R}^{mn}\tilde{R}^{pq}h^{4}$. \ Following
Ref. \cite{randall}, we asumme that matter has only nonzero components on
the 4-dimensional manifold $\Sigma _{4}$, so that we postulate that the
components of the $h^{a}$-field are given by%
\begin{equation}
h^{m}(\phi ,\tilde{x})=e^{g(\phi )}\tilde{h}^{m}(\tilde{x}),\text{ \ }%
h^{4}=0,  \label{o}
\end{equation}%
where $g(\phi )$ is a well behaved function in $0\leqslant \phi <2\pi $. \
This means that the quadratic term $\tilde{\varepsilon}_{mnpq}\tilde{R}^{mn}%
\tilde{R}^{pq}h^{4}$ vanishes.

By replacing (\ref{8t}), (\ref{9t}) and (\ref{10t}) in (\ref{4t}) we find
that the action takes the form%
\begin{eqnarray}
\tilde{S}[\tilde{e},\tilde{h}] &=&\int_{\Sigma _{4}}\tilde{\varepsilon}%
_{mnpq}\left( A\tilde{R}^{mn}\tilde{e}^{p}\tilde{e}^{q}+B\ \tilde{e}^{m}%
\tilde{e}^{n}\tilde{e}^{p}\tilde{e}^{q}+\right.  \notag \\
&&\left. C\tilde{R}^{mn}\tilde{e}^{p}\tilde{h}^{q}+E\tilde{e}^{m}\tilde{e}%
^{n}\tilde{e}^{p}\tilde{h}^{q}\right) ,  \label{999}
\end{eqnarray}%
where%
\begin{equation}
A=2r_{c}\int_{0}^{2\pi }d\phi e^{2f}\left[ \alpha _{3}-\frac{\alpha _{1}l^{2}%
}{r_{c}^{2}}\left( 3f^{\prime 2}+2f^{\prime \prime }\right) \right] ,
\label{12t}
\end{equation}%
\begin{equation}
B=-\frac{1}{r_{c}}\int_{0}^{2\pi }d\phi e^{4f}\left[ \frac{2\alpha _{3}}{3}%
\left( 5f^{\prime 2}+2f^{\prime \prime }\right) -\frac{\alpha _{1}l^{2}}{%
r_{c}^{2}}f^{\prime 2}\left( 5f^{\prime 2}+4f^{\prime \prime }\right) \right]
,  \label{13t}
\end{equation}%
\begin{equation}
C=-\frac{4\alpha _{3}l^{2}}{r_{c}}\int_{0}^{2\pi }d\phi e^{f}e^{g}\left(
f^{\prime 2}+f^{\prime \prime }\right) ,  \label{14t}
\end{equation}%
\begin{equation}
E=\frac{4\alpha _{3}l^{2}}{r_{c}^{3}}\int_{0}^{2\pi }d\phi
e^{3f}e^{g}f^{\prime 2}\left( f^{\prime 2}+f^{\prime \prime }\right) .
\label{15t}
\end{equation}

Since $f(\phi )$ and $g(\phi )$ are arbitrary and continuously
differentiable functions, and since we are working with a cylindrical
variety, if we choose (non-unique choice) $f(\phi )=g\left( \phi \right)
=\ln \left( K+\sin \phi \right) $ with $K=constant>1$, we find that (\ref%
{12t}), (\ref{13t}), (\ref{14t}) and (\ref{15t}) lead to

\begin{equation}
A=\frac{2\pi }{r_{c}}\left[ \alpha _{3}r_{c}^{2}\left( 2K^{2}+1\right)
+\alpha _{1}l^{2}\right] ,  \label{20t}
\end{equation}%
\begin{equation}
B=\frac{\pi }{2r_{c}}\left[ \alpha _{3}\left( 4K^{2}+1\right) -\frac{\alpha
_{1}l^{2}}{2r_{c}^{2}}\right] ,  \label{21t}
\end{equation}

\begin{equation}
C=-4r_{c}^{2}E=\frac{4\pi \alpha _{3}l^{2}}{r_{c}}.  \label{75t}
\end{equation}

Taking into account that $L_{ChS}^{(5)}[e,h]$ flows into $L_{EH}^{(5)}$ when 
$l\longrightarrow 0$ \cite{salg1}, we have that action (\ref{999}) should
lead to the action of Einstein-Hilbert when $l\longrightarrow 0$. From (\ref%
{999}) it is direct to see that this occurs when $A=-1/2$ and $B=\Lambda /12$%
, where $\Lambda $ is the cosmological constant. In this case, from
equations (\ref{20t}), (\ref{21t}) and (\ref{75t}), it is direct to see that

\begin{equation}
\alpha _{1}=-\frac{r_{c}}{4\pi l^{2}}\left[ 1+\frac{\left( 4r_{c}^{2}\Lambda
-3\right) \left( 2K^{2}+1\right) }{3\left( 10K^{2}+3\right) }\right] ,
\label{24t}
\end{equation}%
\begin{equation}
\alpha _{3}=\frac{1}{12\pi r_{c}}\frac{\left( 4r_{c}^{2}\Lambda -3\right) }{%
\left( 10K^{2}+3\right) },  \label{25t}
\end{equation}

\begin{equation}
C=-4r_{c}^{2}E=\frac{l^{2}}{3r_{c}^{2}}\frac{\left( 4r_{c}^{2}\Lambda
-3\right) }{\left( 10K^{2}+3\right) },  \label{27t}
\end{equation}%
and, therefore the action (\ref{999}) takes the form 
\begin{eqnarray}
\tilde{S}[\tilde{e},\tilde{h}] &=&\int_{\Sigma _{4}}\tilde{\varepsilon}%
_{mnpq}\left( -\frac{1}{2}\tilde{R}^{mn}\tilde{e}^{p}\tilde{e}^{q}+\frac{%
\Lambda }{12}\ \tilde{e}^{m}\tilde{e}^{n}\tilde{e}^{p}\tilde{e}^{q}\right. 
\notag \\
&&\left. +C\tilde{R}^{mn}\tilde{e}^{p}\tilde{h}^{q}-\frac{C}{4r_{c}^{2}}%
\tilde{e}^{m}\tilde{e}^{n}\tilde{e}^{p}\tilde{h}^{q}\right) ,  \label{28t'}
\end{eqnarray}%
corresponding to a $4$-dimensional brane embedded in the $5$-dimensional
space-time of the Einstein-Chern-Simons gravity. We can see that, when $%
l\rightarrow 0$ then $C\rightarrow 0$, and hence (\ref{28t'}) becomes the
4-dimensional Einstein-Hilbert action with cosmological constant.

\section{\textbf{An application to cosmology}}

In this section, we will find De Sitter cosmological solutions associated
with the action (\ref{28t'}). We start our analysis by writing the action (%
\ref{28t'}) in tensorial language. \ The first two terms of the Lagrangian (%
\ref{28t'}) can be written as%
\begin{eqnarray}
\tilde{\varepsilon}_{mnpq}\tilde{R}^{mn}\tilde{e}^{p}\tilde{e}^{q} &=&-2%
\sqrt{-\tilde{g}}\tilde{R}d^{4}\tilde{x},  \notag \\
\tilde{\varepsilon}_{mnpq}\tilde{e}^{m}\tilde{e}^{n}\tilde{e}^{p}\tilde{e}%
^{q} &=&-24\sqrt{-\tilde{g}}d^{4}\tilde{x},  \label{29t}
\end{eqnarray}%
where $\tilde{g}$ is the determinant of the $4$-dimensional metric tensor $%
\tilde{g}_{\mu \nu }$, and $\tilde{R}$ is the 4-dimensional Ricci scalar.
The two remaining terms are obtained after some calculations. The results
are given by%
\begin{eqnarray}
\tilde{\varepsilon}_{mnpq}\tilde{R}^{mn}\tilde{e}^{p}\tilde{h}^{q} &=&2\sqrt{%
-\tilde{g}}\left( \tilde{R}\tilde{h}-2\tilde{R}_{\text{ }\nu }^{\mu }\tilde{h%
}_{\text{ }\mu }^{\nu }\right) d^{4}\tilde{x},  \notag \\
\tilde{\varepsilon}_{mnpq}\tilde{e}^{m}\tilde{e}^{n}\tilde{e}^{p}\tilde{h}%
^{q} &=&6\sqrt{-\tilde{g}}\tilde{h}d^{4}\tilde{x},  \label{30t}
\end{eqnarray}%
where $\tilde{h}^{m}=\tilde{h}_{\text{ \ }\mu }^{m}d\tilde{x}^{\mu }$ and $%
\tilde{h}=\tilde{h}_{\text{ }\mu }^{\mu }$. Introducing (\ref{29t}, \ref{30t}%
) in the Lagrangian (\ref{28t'}) we obtain

\begin{equation}
\tilde{S}[\tilde{g},\tilde{h}]=\int d^{4}\tilde{x}\sqrt{-\tilde{g}}\left[
\left( \tilde{R}-2\Lambda \right) +2C\left( \tilde{R}\tilde{h}-2\tilde{R}_{%
\text{ }\nu }^{\mu }\tilde{h}_{\text{ }\mu }^{\nu }\right) +6E\tilde{h}%
\right] .  \label{31t}
\end{equation}

If we consider a maximally symmetric space-time (for instance, the de
Sitter's space), the equation 13.4.6 of Ref. \cite{weinberg} allows us to
write

\begin{equation}
\tilde{h}_{\mu \nu }=\frac{\tilde{F}(\tilde{\varphi})}{4}\tilde{g}_{\mu \nu
},  \label{66t}
\end{equation}%
where $\tilde{F}$ is an arbitrary function of the 4-scalar field $\tilde{%
\varphi}=$ $\tilde{\varphi}(\tilde{x})$.\ This means

\begin{equation}
\tilde{R}_{\text{ }\nu }^{\mu }\tilde{h}_{\text{ }\mu }^{\nu }=\frac{\tilde{F%
}(\tilde{\varphi})}{4}\tilde{R},\text{\ \ \ \ }\tilde{h}=\tilde{h}_{\mu \nu }%
\tilde{g}_{\text{ }}^{\mu \nu }=\tilde{F}(\tilde{\varphi}),
\end{equation}%
so that the action (\ref{31t}) takes the form%
\begin{equation}
\tilde{S}[\tilde{g},\tilde{\varphi}]=\int d^{4}\tilde{x}\sqrt{-\tilde{g}}%
\left[ \left( \tilde{R}-2\Lambda \right) +C\tilde{R}\tilde{F}(\tilde{\varphi}%
)+6E\tilde{F}(\tilde{\varphi})\right] ,  \label{32t}
\end{equation}%
which corresponds to an action for the 4-dimensional gravity coupled
non-minimally to a scalar field. Note that this action has the form%
\begin{equation}
\tilde{S}_{B}=\tilde{S}_{g}+\tilde{S}_{g\varphi }+\tilde{S}_{\varphi },
\end{equation}%
where, $\tilde{S}_{g}$ is a pure gravitational action term, $\tilde{S}%
_{g\varphi }$ is a non-minimal interaction term between gravity and a scalar
field, and $\tilde{S}_{\varphi }$ represents a kind of scalar field
potential. In order to write down the action in the usual way, we define the
constant $\varepsilon $ and the potential $V(\varphi )$ as (removing the
symbols $\sim $ in (\ref{32t}) 
\begin{equation}
\varepsilon =-\frac{\kappa C}{3E}\text{ ,\ \ }V(\varphi )=-\frac{3E}{\kappa }%
F(\varphi ),  \label{100t}
\end{equation}%
where $\kappa $ is the gravitational constant. This permits to rewrite the
action for a $4$-dimensional brane non-minimally coupled to a scalar field,
immersed in a cylindrical 5-dimensional space-time as%
\begin{equation}
S[g,\varphi ]=\int d^{4}x\sqrt{-g}\left\{ \left( R-2\Lambda \right)
+\varepsilon RV(\varphi )-2\kappa V(\varphi )\right\} .  \label{33t}
\end{equation}

The corresponding field equations are obtained by varying the action (\ref%
{33t}) and equalizing it to zero. In fact%
\begin{eqnarray}
\delta \mathcal{L} &\mathcal{=}&\sqrt{-g}\left[ G_{\mu \nu }\left(
1+\varepsilon V\right) +\Lambda g_{\mu \nu }+\kappa g_{\mu \nu
}V+\varepsilon H_{\mu \nu }\right] \delta g^{\mu \nu }  \notag \\
&&\text{ \ \ \ \ }-2\kappa \sqrt{-g}\frac{\partial V}{\partial \varphi }%
\left( 1-\frac{\varepsilon R}{2\kappa }\right) \delta \varphi =0,
\end{eqnarray}%
where 
\begin{equation}
H_{\mu \nu }=g_{\mu \nu }\nabla ^{\lambda }\nabla _{\lambda }V-\nabla _{\mu
}\nabla _{\nu }V
\end{equation}%
and no surface terms have been included. This means that equations
describing the behavior of the $4$-dimensional brane in the presence of the
scalar field $\varphi $ are given by%
\begin{equation}
G_{\mu \nu }\left( 1+\varepsilon V\right) +\Lambda g_{\mu \nu }+\varepsilon
H_{\mu \nu }=-\kappa g_{\mu \nu }V,  \label{z1}
\end{equation}%
\begin{equation}
\frac{\partial V}{\partial \varphi }\left( 1-\frac{\varepsilon R}{2\kappa }%
\right) =0.  \label{z2}
\end{equation}

Note that if $\varepsilon \rightarrow 0$, then the scalar field potential
behaves as a constant $V_{0}$, and the Einstein's vacuum field equations
with (a modified) cosmological constant are recovered:%
\begin{equation}
G_{\mu \nu }+\Lambda _{0}g_{\mu \nu }=0,
\end{equation}%
where $\Lambda _{0}=\left( \Lambda +\kappa V_{0}\right) /\left(
1+\varepsilon V_{0}\right) \approx \left( \Lambda +\kappa V_{0}\right) .$

In order to construct a model of universe based on equations (\ref{z1})-(\ref%
{z2}), we consider the Friedmann-Lemaitre-Robertson-Walker metric%
\begin{equation}
ds^{2}=-dt^{2}+a(t)^{2}\left[ \frac{dr^{2}}{1-kr^{2}}+r^{2}\left( d\theta
^{2}+\sin ^{2}\theta d\psi ^{2}\right) \right] ,
\end{equation}%
where $a(t)$ is the so called "scale factor" and $k=0,+1,-1$ describes flat,
spherical and hyperbolic spatial geometries, respectively. Following the
usual procedure, we find that the Friedmann-Lemaitre-Robertson-Walker type
equations, are given by%
\begin{equation}
3\frac{\dot{a}^{2}+k}{a^{2}}\left( 1+\varepsilon V\right) -\Lambda
+3\varepsilon \frac{\dot{a}}{a}\dot{\varphi}\frac{\partial V}{\partial
\varphi }=\kappa V,  \label{ax}
\end{equation}%
\begin{equation}
\left( 2\frac{\ddot{a}}{a}+\frac{\dot{a}^{2}+k}{a^{2}}\right) \left(
1+\varepsilon V\right) -\Lambda +\varepsilon \left[ \dot{\varphi}^{2}\frac{%
\partial ^{2}V}{\partial \varphi ^{2}}+\left( \ddot{\varphi}+2\frac{\dot{a}}{%
a}\dot{\varphi}\right) \frac{\partial V}{\partial \varphi }\right] =\kappa V,
\label{cx}
\end{equation}%
\begin{equation}
\frac{\partial V}{\partial \varphi }\left[ 1-\frac{3\varepsilon }{\kappa }%
\left( \frac{\ddot{a}}{a}+\frac{\dot{a}^{2}+k}{a^{2}}\right) \right] =0.
\label{bx}
\end{equation}

Assuming that $\partial V/\partial \varphi \neq 0$, equation (\ref{bx}) has
the solutions%
\begin{equation}
a(t)=\left \{ 
\begin{array}{c}
\sqrt{\frac{6\varepsilon }{\kappa }}\sinh \left( \sqrt{\frac{\kappa }{%
6\varepsilon }}t\right) ,\text{ \ \ \ }k=-1; \\ 
C_{1}\exp \left( \sqrt{\frac{\kappa }{6\varepsilon }}t\right) ,\text{ \ \ \ }%
k=0; \\ 
\sqrt{\frac{6\varepsilon }{\kappa }}\cosh \left( \sqrt{\frac{\kappa }{%
6\varepsilon }}t\right) ,\text{ \ \ \ }k=+1;%
\end{array}%
\right. ,  \label{t}
\end{equation}%
representing the maximally symmetric De Sitter space-time. Here, $%
\varepsilon >0$ and $C_{1}$ is an arbitrary constant, which can be chosen
equal to $1$.

In the flat case $k=0$, equations (\ref{ax}) and (\ref{cx}) can be solved
exactly for $\varphi $. Replacing $a=e^{\sqrt{\kappa /6\varepsilon }t}$ in (%
\ref{ax}), and assuming that $V$ has the form%
\begin{equation}
V(\varphi )=\lambda \varphi ^{n},  \label{61x}
\end{equation}%
where $\lambda $ and $n$ \ are constants, we obtain the following equation%
\begin{equation}
\left( \frac{\kappa }{2\varepsilon }-\Lambda \right) +\sqrt{\frac{\kappa }{2}%
}V\left[ \sqrt{3\varepsilon }n\frac{\dot{\varphi}}{\varphi }-\sqrt{\frac{%
\kappa }{2}}\right] =0,
\end{equation}%
which has as solution%
\begin{equation}
\Lambda =\frac{\kappa }{2\varepsilon }\text{ ,\ \ \ \ \ \ }\frac{\dot{\varphi%
}}{\varphi }=\frac{1}{n}\sqrt{\frac{\kappa }{6\varepsilon }}=\frac{1}{n}%
\sqrt{\frac{\Lambda }{3}},  \label{62x}
\end{equation}%
and therefore%
\begin{equation}
\varphi (t)=C_{2}e^{(1/n)\sqrt{\Lambda /3}t},  \label{c}
\end{equation}%
where $C_{2}$ is an arbitrary constant which can be chosen equal to $1$.

On the other hand, if we replace $a=e^{\sqrt{\kappa /6\varepsilon }t}$, (\ref%
{61x}), (\ref{62x}) and (\ref{c}) in equation (\ref{cx}), we see that the
equality $0=0$ is satisfied, which means that no constraints are imposed on
the constants $n$ and $\lambda $ in the scalar field potential $V$.

Consequently, the solution of the system of equations (\ref{ax})-(\ref{bx})
is given by%
\begin{equation}
a(t)=e^{\sqrt{\Lambda /3}t},\text{ \ \ \ \ \ }\varphi (t)=e^{(1/n)\sqrt{%
\Lambda /3}t}\text{.}  \label{d}
\end{equation}

This solution describes an accelerated flat expanding universe in the
presence of a scalar field that exhibits an exponential behavior.

Observe that solution (\ref{c}) is also valid for late time cosmology in the
cases with spherical and hyperbolic spatial geometries, namely $k=\pm 1$,
where $\frac{\dot{a}}{a}\approx \sqrt{\frac{\kappa }{6\varepsilon }}=\sqrt{%
\frac{\Lambda }{3}}$.

\section{\textbf{Concluding remarks}}

In this work, we have obtained an action and its corresponding field
equations, for a $4$-dimensional brane embedded in the $5$-dimensional
space-time of the Einstein-Chern-Simons theory of gravity. This framework\
leads to the $4$-dimensional general relativity with cosmological constant
in the limit $l\rightarrow 0$. We have also obtained an interpretation of
the $h^{a}$ matter field present in the Einstein-Chern-Simons gravity
action, which was related to a scalar field. As an application, the
Friedmann-Lemaitre-Robertson-Walker equations for cosmology were found and
some De Sitter accelerated solutions were obtained, considering a scalar
field potential with the form $V=\lambda \varphi ^{n}$, where $\lambda $ and 
$n$ represent arbitrary constants.

In Ref. \cite{salg4} it was found that the Lagrangian (\ref{1t}) can be
written in the form (\ref{ocho}) (see appendix). From (\ref{ocho}) we can
see that kinetic terms corresponding to the fields $h^{a}$ and $k^{ab}$,
absent in the Lagrangian (\ref{1t}), are present in the surface term of the
Lagrangian (\ref{ocho}) through equation (\ref{8'}). This situation is
common to all Chern-Simons theories. This has the consequence that the
action (\ref{33t}) does not have the kinetic term for the scalar field $%
\varphi $.

It could be interesting to add a kinetic term to the $4$-dimensional brane
action. In this case, the action (\ref{33t}) takes the form%
\begin{equation}
S[g,\varphi ]=\int d^{4}x\sqrt{-g}\left\{ \left( R-2\Lambda \right)
+\varepsilon RV(\varphi )-2\kappa \left[ \frac{1}{2}\left( \nabla _{\mu
}\varphi \right) \left( \nabla ^{\mu }\varphi \right) +V(\varphi )\right]
\right\} .  \label{34t}
\end{equation}

The corresponding field equations are given by

\begin{equation}
G_{\mu \nu }\left( 1+\varepsilon V\right) +\Lambda g_{\mu \nu }+\varepsilon
H_{\mu \nu }=\kappa T_{\mu \nu }^{\varphi },  \label{40t}
\end{equation}%
\begin{equation}
\nabla _{\mu }\nabla ^{\mu }\varphi -\frac{\partial V}{\partial \varphi }%
\left( 1-\frac{\varepsilon R}{2\kappa }\right) =0,  \label{41t}
\end{equation}%
where $T_{\mu \nu }^{\varphi }$ is the energy-momentum tensor of the scalar
field%
\begin{equation}
T_{\mu \nu }^{\varphi }=\nabla _{\mu }\varphi \nabla _{\nu }\varphi -g_{\mu
\nu }\left( \frac{1}{2}\nabla ^{\lambda }\varphi \nabla _{\lambda }\varphi
+V\right) ,  \label{37t}
\end{equation}%
and the rank-2 tensor $H_{\mu \nu }$ is defined as%
\begin{equation}
H_{\mu \nu }=g_{\mu \nu }\nabla ^{\lambda }\nabla _{\lambda }V-\nabla _{\mu
}\nabla _{\nu }V.  \label{38t}
\end{equation}

Note that if $\varepsilon \rightarrow 0$, then the Einstein field equations
with cosmological constant, in the presence of a scalar field, are recovered,%
\begin{equation}
G_{\mu \nu }+\Lambda g_{\mu \nu }=\kappa T_{\mu \nu }^{\varphi },\text{ \ \ }%
\nabla _{\mu }\nabla ^{\mu }\varphi -\frac{\partial V}{\partial \varphi }=0.
\end{equation}

Following the usual procedure, we find that the
Friedmann-Lemaitre-Robertson-Walker type equations are%
\begin{equation}
3\left( \frac{\dot{a}^{2}+k}{a^{2}}\right) \left( 1+\varepsilon V\right)
-\Lambda +3\varepsilon \frac{\dot{a}}{a}\dot{\varphi}\frac{\partial V}{%
\partial \varphi }=\kappa \left( \frac{1}{2}\dot{\varphi}^{2}+V\right) ,
\label{48x}
\end{equation}%
\begin{equation}
\left( 2\frac{\ddot{a}}{a}+\frac{\dot{a}^{2}+k}{a^{2}}\right) \left(
1+\varepsilon V\right) -\Lambda +\varepsilon \left[ \dot{\varphi}^{2}\frac{%
\partial ^{2}V}{\partial \varphi ^{2}}+\left( \ddot{\varphi}+2\frac{\dot{a}}{%
a}\dot{\varphi}\right) \frac{\partial V}{\partial \varphi }\right] =-\kappa
\left( \frac{1}{2}\dot{\varphi}^{2}-V\right) ,  \label{49x}
\end{equation}%
\begin{equation}
\ddot{\varphi}+3\frac{\dot{a}}{a}\dot{\varphi}+\frac{\partial V}{\partial
\varphi }\left[ 1-\frac{3\varepsilon }{\kappa }\left( \frac{\ddot{a}}{a}+%
\frac{\dot{a}^{2}+k}{a^{2}}\right) \right] =0.  \label{50x}
\end{equation}

From here, we can see that in the limit $\varepsilon \rightarrow 0$, the
Friedmann equations with a scalar field are recovered (see e.g. Ref. \cite%
{garcia}):%
\begin{equation}
3\left( \frac{\dot{a}^{2}+k}{a^{2}}\right) -\Lambda =\kappa \left( \frac{1}{2%
}\dot{\varphi}^{2}+V\right) ,
\end{equation}%
\begin{equation}
2\frac{\ddot{a}}{a}+\frac{\dot{a}^{2}+k}{a^{2}}-\Lambda =-\kappa \left( 
\frac{1}{2}\dot{\varphi}^{2}-V\right) ,
\end{equation}%
\begin{equation}
\ddot{\varphi}+3\frac{\dot{a}}{a}\dot{\varphi}+\frac{\partial V}{\partial
\varphi }=0.
\end{equation}

In the case of a constant scalar field $\varphi =\varphi _{0}$, we have that 
$\ddot{\varphi}=\dot{\varphi}=0$ and $V(\varphi _{0})=V_{0}$. Then, we find
the following solution for the system (\ref{48x})-(\ref{50x}):

\begin{equation}
a(t)=\left\{ 
\begin{array}{c}
\sqrt{\frac{6\varepsilon }{\kappa }}\sinh \left( \sqrt{\frac{\kappa }{%
6\varepsilon }}t\right) ,\text{ \ \ \ }k=-1, \\ 
C\exp \left( \sqrt{\frac{\kappa }{6\varepsilon }}t\right) ,\text{ \ \ \ }k=0,
\\ 
\sqrt{\frac{6\varepsilon }{\kappa }}\cosh \left( \sqrt{\frac{\kappa }{%
6\varepsilon }}t\right) ,\text{ \ \ \ }k=+1,%
\end{array}%
\right.  \label{86x}
\end{equation}%
where $C$ is an arbitrary constant, $\varepsilon >0$ and the constant scalar
field potential takes the form 
\begin{equation}
V_{0}=\frac{\kappa -2\varepsilon \Lambda }{\kappa \varepsilon }.
\end{equation}

If we wish to interpret the scalar field $\varphi $ as a Klein-Gordon
particle, with potential $V\left( \varphi \right) =\left( m^{2}/2\right)
\varphi ^{2}$, then the mass of this field has to be given by%
\begin{equation}
m=\frac{1}{\left\vert \varphi _{0}\right\vert }\sqrt{\frac{2\left( \kappa
-2\varepsilon \Lambda \right) }{\kappa \varepsilon }},  \label{m}
\end{equation}%
where $\kappa -2\varepsilon \Lambda >0.$

On the another hand, in the case that $k=0$ and the scalar field is
non-constant, equations (\ref{48x})-(\ref{50x}) take the form%
\begin{equation}
3\frac{\dot{a}^{2}}{a^{2}}\left( 1+\varepsilon V\right) -\Lambda
+3\varepsilon \frac{\dot{a}}{a}\dot{\varphi}\frac{\partial V}{\partial
\varphi }=\kappa \left( \frac{1}{2}\dot{\varphi}^{2}+V\right) ,  \label{51x}
\end{equation}%
\begin{equation}
\left( 2\frac{\ddot{a}}{a}+\frac{\dot{a}^{2}}{a^{2}}\right) \left(
1+\varepsilon V\right) -\Lambda +\varepsilon \left[ \dot{\varphi}^{2}\frac{%
\partial ^{2}V}{\partial \varphi ^{2}}+\left( \ddot{\varphi}+2\frac{\dot{a}}{%
a}\dot{\varphi}\right) \frac{\partial V}{\partial \varphi }\right] =-\kappa
\left( \frac{1}{2}\dot{\varphi}^{2}-V\right) ,  \label{52x}
\end{equation}%
\begin{equation}
\ddot{\varphi}+3\frac{\dot{a}}{a}\dot{\varphi}+\frac{\partial V}{\partial
\varphi }\left[ 1-\frac{3\varepsilon }{\kappa }\left( \frac{\ddot{a}}{a}+%
\frac{\dot{a}^{2}}{a^{2}}\right) \right] =0.  \label{53x}
\end{equation}

A solution for this system can be found by considering first%
\begin{equation}
1-\frac{3\varepsilon }{\kappa }\left( \frac{\ddot{a}}{a}+\frac{\dot{a}^{2}}{%
a^{2}}\right) =0  \label{54x}
\end{equation}%
in equation (\ref{53x}). The solution of (\ref{54x}) for the scale factor is
given by 
\begin{equation}
a(t)=C_{1}e^{\sqrt{\kappa /6\varepsilon }t},  \label{x54}
\end{equation}%
where $C_{1}$ is an arbitrary constant and $\varepsilon >0$.

Conversely, the replacement of (\ref{54x}) and (\ref{x54}) in (\ref{53x}),
leads to the equation for the scalar field%
\begin{equation}
\ddot{\varphi}+\frac{\kappa }{2\varepsilon }\dot{\varphi}=0,
\end{equation}%
whose solution is%
\begin{equation}
\varphi (t)=C_{2}e^{-\sqrt{3\kappa /2\varepsilon }t}+C_{3},
\end{equation}%
where $C_{2}$ and $C_{3}$ are arbitrary constants.

Choosing $C_{1}=C_{2}=1$ and $C_{3}=0$, we arrive at the following solution
for equation (\ref{53x}),%
\begin{equation}
a(t)=e^{\sqrt{\kappa /6\varepsilon }t},\text{ \ \ \ \ \ }\varphi (t)=e^{-%
\sqrt{3\kappa /2\varepsilon }t},  \label{55x}
\end{equation}%
which describes a De Sitter universe in exponential accelerated expansion,
in presence of a scalar field that exhibits an exponential decay.

For consistency, the solution (\ref{55x}) must also be a solution of the
equations (\ref{51x}) and (\ref{52x}). By replacing (\ref{55x}) in (\ref{51x}%
), and assuming that the potential of the scalar field is given by%
\begin{equation}
V(\varphi )=\lambda \varphi ^{n},  \label{56x}
\end{equation}%
with $\lambda $ and $n$ constants, we obtain the following equation%
\begin{equation}
\left( \frac{\kappa }{2\varepsilon }-\Lambda \right) -\frac{\kappa }{2}\left[
\frac{3\kappa }{2\varepsilon }\varphi ^{2}+\left( 1+3n\right) V\right] =0,
\end{equation}%
which is satisfied choosing the cosmological constant and the potential as
follows%
\begin{equation}
\Lambda =\frac{\kappa }{2\varepsilon }\text{ ,\ \ \ \ \ }V(\varphi )=-\frac{%
3\kappa }{2\varepsilon \left( 1+3n\right) }\varphi ^{2}=-\frac{3\Lambda }{%
\left( 1+3n\right) }\varphi ^{2}.  \label{63x}
\end{equation}

According to equations (\ref{56x}) and (\ref{63x}), we see that%
\begin{equation}
n=2,\text{ \ \ \ \ \ }\lambda =-\frac{3\Lambda }{7}\text{.}
\end{equation}

Observe that $\Lambda =\kappa /2\varepsilon >0,$ then $\lambda <0$.
Therefore, the scalar field $\varphi $, with potential%
\begin{equation}
V(\varphi )=\lambda \varphi ^{n}=-\frac{3\Lambda }{7}\varphi ^{2}=\frac{m^{2}%
}{2}\varphi ^{2},
\end{equation}%
might be identified with a tachyon, a hypothetical particle with imaginary
mass, whose speed is higher than the speed of light. In this case, the mass
of the particle is%
\begin{equation}
m=i\sqrt{\frac{6\Lambda }{7}}.  \label{m2}
\end{equation}

Note that equation (\ref{52x}) is consistant with the results found. The
solution (\ref{55x}) can be rewritten in terms of the cosmological constant
as%
\begin{equation}
a(t)=e^{\sqrt{\Lambda /3}t},\text{ \ \ \ \ }\varphi (t)=e^{-\sqrt{3\Lambda }%
t}\text{.}  \label{s}
\end{equation}

Finally, it is worth noting that in Ref. \cite{edelstein} a 5-dimensional
Chern-Simons action for gravity, invariant under the Poincar\'{e} group
enlarged by an Abelian ideal, was found. Equations and solutions found in
the present article, obtained from Einstein-Chern-Simons gravity \cite{salg1}%
, can be easily replicated for the theory in Ref. \cite{edelstein}, through
the replacement $\alpha _{1}=0$ and $\alpha _{3}=l=1$.

It might be of interest to note some similarities and differences between
the four-dimensional Lagrangian of the action (\ref{28t'}) and the
Lagrangian of the gravity theory based on the Maxwell algebra in four
dimensions proposed by Azcarraga, Kamimura and Lukierski in Ref. \cite%
{azcarraga}\textit{. \ }

\textit{The }Lagrangian of the action (\ref{28t'}) is obtained from the so
called Einstein-Chern-Simons Lagrangian (\ref{ocho}) for the $\mathfrak{B}$
Lie algebra (\ref{cuatro}), whose generators are given by $\left\{
J_{ab},P_{a},Z_{ab},Z_{a}\right\} $ \cite{salg1}.

The gravity theory of Ref. \cite{azcarraga} is based on the Maxwell algebra
whose generators are given by $\left\{ J_{ab},P_{a},Z_{ab}\right\} .$ This
algebra can also be obtained from the anti de Sitter ($AdS$) algebra via the
mentioned S-expansion procedure.

The corresponding one-form gauge connection and the two-form curvature are
given by%
\begin{eqnarray}
\boldsymbol{A} &=&e^{a}\boldsymbol{P}_{a}+\frac{1}{2}\omega ^{ab}\boldsymbol{%
J}_{ab}+\frac{1}{2}k^{ab}\boldsymbol{Z}_{ab},  \notag \\
\boldsymbol{F} &=&T^{a}\boldsymbol{P}_{a}+\frac{1}{2}R^{ab}\boldsymbol{J}%
_{ab}+\frac{1}{2}\mathcal{F}^{ab}\boldsymbol{Z}_{ab},  \label{resp4}
\end{eqnarray}%
where $T^{a}=de^{a}+\omega _{\text{ \ }c}^{a}e^{c}=\mathrm{D}_{\omega }e^{a}$%
, $R^{ab}=d\omega ^{ab}+\omega _{\text{ \ }c}^{a}\omega ^{cb},$ $\mathcal{F}%
^{ab}=\mathrm{D}_{\omega }k^{ab}+\Lambda e^{a}e^{b},$ which allows us to
construct the Lagrangians ($29)$ of the Ref. \cite{azcarraga}:%
\begin{eqnarray}
\mathcal{L} &=&-\frac{1}{2\kappa }\varepsilon _{abcd}R^{ab}e^{c}e^{d}+\frac{%
\lambda }{4\kappa }\varepsilon _{abcd}e^{a}e^{b}e^{c}e^{d}  \notag \\
&&+\frac{\mu }{2\kappa }\varepsilon _{abcd}\left( DA\right) ^{ab}e^{c}e^{d}+%
\frac{\mu ^{2}}{4\kappa \lambda }\varepsilon _{abcd}\left( DA\right)
^{ab}\left( DA\right) ^{cd},  \label{29}
\end{eqnarray}%
\textit{where }$\mathcal{F}^{ab}\equiv F^{ab}$ and $k^{ab}\equiv A^{ab}$.

Since Maxwell's algebra is contained in algebra $\mathfrak{B}$, it would be
reasonable to expect that the action (\ref{29}) should be a particular case
of the action (\ref{28t'}). \ However, this fact is not observed when the
corresponding lagrangians are compared. \ The reason for this discrepancy is
due to the fact that we have imposed the freedom of torsion condition on
Lagrangian (\ref{1t}). The imposition of this condition eliminates from the
Lagrangian (\ref{1t}) the field $k^{ab}\equiv A^{ab}$. Comparison of the
Lagrangians (\ref{28t'}) and (\ref{29}) shows that they have in common only
the Einstein-Hilbert and the cosmological terms. An interesting problem
could be consider equation (\ref{1t}) without imposing the torsion-free
condition, carry out the same procedure that leads to a Lagrangian analogous
to Lagrangian (\ref{28t'}) and then comparing it with the Lagrangian (\ref%
{29}).

It may be also of interest to note that although the two Lagrangians contain
a cosmological term, they have different origins. The cosmological term of
the Lagrangian (\ref{29}) has its origin in the commutation relation $\left[
P_{a},P_{b}\right] =Z_{ab}$, this means in the field $k^{ab}$, while the
cosmological term of the Lagrangian (\ref{28t'}) appears when the $5$%
-dimensional Randall-Sundrum metric is introduced.

It might be of interest to note that the fields $k^{ab}$, associated with
the generators $Z_{ab}$, plays a role similar to that played by the field $%
A^{ab}$ in Ref. \cite{azcarraga}, that is, to introduce a cosmological term
in an alternative way that generalizes standard gravity. However, as we have
already mentioned, this term is not present in our approach. \ On the other
hand, the $h^{a}$ field, associated with the generators $Z_{a},$ plays the
role of a scalar field in the case of maximally symmetric space-times.

\section{\textbf{Appendix}}

A ChS lagrangian in $d=2n+1$ dimensions is defined to be the following local
function of a one-form gauge connection $\boldsymbol{A}$: 
\begin{eqnarray}
L_{\text{ }ChS}^{\left( 2n+1\right) }\left( \boldsymbol{A}\right) &=&\left(
n+1\right) k\int_{0}^{1}dt\left\langle \boldsymbol{A}\left( t\text{d}%
\boldsymbol{A}+t^{2}\boldsymbol{A}^{2}\right) ^{n}\right\rangle ,  \notag \\
&=&\left( n+1\right) k\int_{0}^{1}dt\left\langle \boldsymbol{A}\left( t\,%
\boldsymbol{F}+(t^{2}-t)\boldsymbol{A}^{2}\right) ^{n}\right\rangle
\label{uno}
\end{eqnarray}%
where $\left\langle \cdots \right\rangle $ denotes a invariant tensor for
the corresponding Lie algebra$,$ $F=dA+AA$ is the corresponding the two-form
curvature and $k$ is a constant \cite{zan}.

Some time ago was shown that the standard, five-di\-men\-sio\-nal General
Relativity can be obtained from Chern-Simons gravity theory for a certain
Lie algebra $\mathfrak{B}$ \cite{salg1}, whose generators $\left\{
J_{ab},P_{a},Z_{ab},Z_{a}\right\} $ satisfy the commutation relationships

\begin{equation*}
\left[ J_{ab},J_{cd}\right] =\eta _{cb}J_{ad}-\eta _{ca}J_{bd}+\eta
_{db}J_{ca}-\eta _{da}J_{cb}
\end{equation*}%
\begin{equation*}
\left[ J_{ab},P_{c}\right] =\eta _{cb}P_{a}-\eta _{ca}P_{b}
\end{equation*}%
\begin{equation*}
\text{\ }\left[ P_{a},P_{b}\right] =Z_{ab}
\end{equation*}%
\begin{equation*}
\left[ J_{ab},Z_{cd}\right] =\eta _{cb}Z_{ad}-\eta _{ca}Z_{bd}+\eta
_{db}Z_{ca}-\eta _{da}Z_{cb}
\end{equation*}%
\begin{equation}
\left[ J_{ab},Z_{c}\right] =\eta _{cb}Z_{a}-\eta _{ca}Z_{b}  \label{cuatro}
\end{equation}%
\begin{equation*}
\left[ Z_{ab},P_{c}\right] =\eta _{cb}Z_{a}-\eta _{ca}Z_{b}
\end{equation*}

This algebra was obtained from the anti de Sitter ($AdS$) algebra and a
particular semigroup $S$ by means of the S-expansion procedure introduced in
Refs. \cite{salg2}, \cite{salg3}.

In order to write down a Chern--Simons lagrangian for the $\mathfrak{B}$
algebra, we start from the one-form gauge connection%
\begin{equation}
\boldsymbol{A}=\frac{1}{2}\omega ^{ab}\boldsymbol{J}_{ab}+\frac{1}{l}e^{a}%
\boldsymbol{P}_{a}+\frac{1}{2}k^{ab}\boldsymbol{Z}_{ab}+\frac{1}{l}h^{a}%
\boldsymbol{Z}_{a},  \label{cinco}
\end{equation}%
and the two-form curvature%
\begin{eqnarray}
\boldsymbol{F} &=&\frac{1}{2}R^{ab}\boldsymbol{J}_{ab}+\frac{1}{l}T^{a}%
\boldsymbol{P}_{a}+\frac{1}{2}\left( \mathrm{D}_{\omega }k^{ab}+\frac{1}{%
l^{2}}e^{a}e^{b}\right) \boldsymbol{Z}_{ab}  \notag \\
&&+\frac{1}{l}\left( \mathrm{D}_{\omega }h^{a}+k_{\text{ \ \ }%
b}^{a}e^{b}\right) \boldsymbol{Z}_{a}.  \label{seis}
\end{eqnarray}

A Chern-Simons lagrangian in $d=5$ dimensions is defined to be the following
local function of a one-form gauge connection $\boldsymbol{A}$: 
\begin{equation}
L_{ChS}^{\left( 5\right) }\left( \boldsymbol{A}\right) =k\left\langle 
\boldsymbol{AF}^{2}-\frac{1}{2}\boldsymbol{A}^{3}\boldsymbol{F+}\frac{1}{10}%
\boldsymbol{A}^{5}\right\rangle ,  \label{lcs}
\end{equation}%
where $\left\langle \cdots \right\rangle $ denotes a invariant tensor for
the corresponding Lie algebra$,$ $F=dA+AA$ is the corresponding the two-form
curvature and $k$ is a constant \cite{zan}.

Using theorem~VII.2 of Ref.~\cite{salg2}, and the extended Cartan's homotopy
formula as in Ref. \cite{irs1}, and integrating by parts, it is possible to
write down the Chern-Simons Lagrangian in five dimensions for the $\mathcal{B%
}$ algebra as \cite{salg1}, \cite{salg4}

\begin{align}
L_{EChS}^{(5)}& =\alpha _{1}l^{2}\varepsilon _{abcde}R^{ab}R^{cd}e^{e} 
\notag \\
& \quad +\alpha _{3}\epsilon _{abcde}\left( \frac{2}{3}%
R^{ab}e^{c}e^{d}e^{e}+2l^{2}k^{ab}R^{cd}T^{e}+l^{2}R^{ab}R^{cd}h^{e}\right) 
\notag \\
& \qquad +dB_{\text{EChS}}^{(4)}  \label{ocho}
\end{align}%
where the surface term $B_{\text{EChS}}^{(4)}$ is given by

\begin{align}
B_{EChS}^{(4)}& =\alpha _{1}l^{2}\epsilon _{abcde}e^{a}\omega ^{bc}\left( 
\frac{2}{3}d\omega ^{de}+\frac{1}{2}\omega _{\text{ \ }f}^{d}\omega
^{fe}\right)  \notag \\
& \quad +\alpha _{3}\epsilon _{abcde}\left[ l^{2}\left( h^{a}\omega
^{bc}+k^{ab}e^{c}\right) \left( \frac{2}{3}d\omega ^{de}+\frac{1}{2}\omega _{%
\text{ \ }f}^{d}\omega ^{fe}\right) \right.  \notag \\
& \qquad \qquad \qquad \left. +l^{2}k^{ab}\omega ^{cd}\left( \frac{2}{3}%
de^{e}+\frac{1}{2}\omega _{\text{ \ }f}^{d}e^{e}\right) +\frac{1}{6}%
e^{a}e^{b}e^{c}\omega ^{de}\right]  \label{8'}
\end{align}%
and where $\alpha _{1}$, $\alpha _{3}$ are parameters of the theory, $l$ is
a coupling constant, $R^{ab}=d\omega ^{ab}+\omega _{b}^{a}$ corresponds to
the curvature $2$-form in the first-order formalism related to the $1$-form
spin connection, and $e^{a}$, $h^{a}$ and $k^{ab}$ are others gauge fields
presents in the theory.

Finally it might be necessary to notice that:

\textbf{(a)} The lagrangian is splitt into two independient pieces, one
proportional to $\alpha _{1}$ and the other to $\alpha _{3}.$ The piece
proportional to $\alpha _{1}$ corresponds to the In\"{o}n\"{u}-Wigner
contraction of the Chern--Simons Lagrangian corresponding to $AdS$-algebra,
and therefore it is the Chern-Simons Lagrangian for the Poincar\'{e}-Lie
algebra $\mathrm{ISO}\left( 4,1\right) $. The piece proportional to $\alpha
_{3}$ contains the Einstein--Hilbert term $\varepsilon
_{abcde}R^{ab}e^{c}e^{d}e^{e}$ plus non-linear couplings between the
curvature and the bosonic "matter" fields $k^{ab}$ and $h^{a}$, where the
parameter $l^{2}$ can be interpreted as a kind of coupling constant.

\textbf{(b)} When the constant $\alpha _{1}$ vanishes, the lagrangian (\ref%
{ocho}) almost exactly matches the one given in Ref.~\cite{edelstein}, the
only difference being that in our case the coupling constant $l^{2}$ appears
explicitly in the last two terms.

The presence or absence of the coupling constant $l$ in the lagrangian could
seem like a minor or trivial matter, but it is not. As the authors of Ref.~%
\cite{edelstein} clearly state, the presence of the Einstein--Hilbert term
in this kind of action does not guarantee that the dynamics will be that of
general relativity. In general, extra constraints on the geometry do appear,
even around a "vacuum" solution with $k^{ab}=h^{a}=0$. In fact, the
variation of the lagrangian, modulo boundary terms, can be written as%
\begin{eqnarray}
\delta L_{\mathrm{CS}}^{\left( 5\right) } &=&\epsilon _{abcde}\left( 2\alpha
_{3}R^{ab}e^{c}e^{d}+\alpha _{1}l^{2}R^{ab}R^{cd}+2\alpha _{3}l^{2}\mathrm{D}%
_{\omega }k^{ab}R^{cd}\right) \delta e^{e}  \notag \\
&&+\alpha _{3}l^{2}\epsilon _{abcde}R^{ab}R^{cd}\delta h^{e}+  \notag \\
&&+2\epsilon _{abcde}\delta \omega ^{ab}\left( \alpha
_{1}l^{2}R^{cd}T^{e}+\alpha _{3}l^{2}\mathrm{D}k^{cd}T^{e}+\alpha
_{3}e^{c}e^{d}T^{e}\right.  \notag \\
&&\left. +\alpha _{3}l^{2}R^{cd}\mathrm{D}h^{e}+\alpha _{3}l^{2}R^{cd}k_{%
\text{ \ }f}^{e}e^{f}+2\alpha _{3}l^{2}\epsilon _{abcde}\delta
k^{ab}R^{cd}T^{e}\right) .
\end{eqnarray}

Therefore, when the condition $\alpha _{1}=0$ is chosen, the torsionless
condition imposed, and a solution without matter $(k^{ab}=h^{a}=0$) is
picked out, we are left with%
\begin{equation}
\delta L_{\mathrm{CS}}^{\left( 5\right) }=2\alpha _{3}\epsilon
_{abcde}R^{ab}e^{c}e^{d}\delta e^{e}+\alpha _{3}l^{2}\varepsilon
_{abcde}R^{ab}R^{cd}\delta h^{e}.
\end{equation}

In this way, besides general relativity equations of motions $\epsilon
_{abcde}R^{ab}e^{c}e^{d}=0,$ the equations of motion of pure Gauss-Bonnet
theory $\varepsilon _{abcde}R^{ab}R^{cd}=0$ do also appear as an anomalous
constraint on the geometry.

It is at this point where the presence of the coupling constant $l$ makes
the difference. In the present approach, it does play the role of a coupling
constant between geometry and \textquotedblleft matter\textquotedblright .
For this reason, in this case the limit $l\rightarrow 0$ leads to the
Einstein--Hilbert term in the lagrangian, 
\begin{equation}
L_{\mathrm{CS}}^{\left( 5\right) }=\frac{2}{3}\alpha _{3}\epsilon
_{abcde}R^{ab}e^{c}e^{d}e^{e}.
\end{equation}

In the same way, when we impose the weak limit of coupling constant, $%
l\rightarrow 0,$ the extra constraints just vanish, and $\delta L_{\mathrm{CS%
}}^{\left( 5\right) }=0$ lead us to just the Einstein--Hilbert dynamics in
the vaccum,%
\begin{equation}
\delta L_{\mathrm{CS}}^{\left( 5\right) }=2\alpha _{3}\epsilon
_{abcde}R^{ab}e^{c}e^{d}\delta e^{e}+2\alpha _{3}\epsilon _{abcde}\delta
\omega ^{ab}e^{c}e^{d}T^{e}.
\end{equation}

It is interesting to observe that the argument given here is not just a
five-dimensional accident; at every odd dimension, it is possible to realize
the $S$-expansion in the way sketched here, to take the weak limit of
coupling constant, $l\rightarrow 0,$ and recover Einstein--Hilbert gravity.

\begin{acknowledgement}
This work was supported in part by Univesidad de los Lagos, Chile. Three of
the authors (RD, FG, MP) were supported by Universidad de Los Lagos Grant \#
R12/18. One of the authors (PS) \textit{was supported in part }by \textit{\
FONDECYT Grants }No.\textit{\ 1180681} from the Government of Chile. The
authors would like to thank Luis Ruiz-Pailalef for the careful reading of
the manuscript.
\end{acknowledgement}


\begin{thebibliography}{99}
\bibitem{kaluza} T. Kaluza, Sitzungsber. Preuss. Akad. Wiss. Berlin (Math.
Phys.) (1921) 966.

\bibitem{klein} O. Klein, Z. Phys. 37 (1926) 895.

\bibitem{randall} L. Randall, R. Sundrum, Phys. Rev. Lett. 83 (17), (1999)
3370.

\bibitem{randall1} L. Randall, R. Sundrum, Phys. Rev. Lett. 83 (17), (1999)
4690.

\bibitem{salg1} F. Izaurieta, P. Minning, \ A. P\'{e}rez, E. Rodr\'{\i}guez,
P. Salgado. Phys. Lett. B 678 (2009) 213.

\bibitem{salg2} F. Izaurieta, E. Rodr\'{\i}guez, P. Salgado, Jour. Math.
Phys. 47 (2006) 123512.

\bibitem{salg3} F. Izaurieta, A. P\'{e}rez, E. Rodr\'{\i}guez and P.
Salgado, J. Math. Phys. 50 (2009) 073511.

\bibitem{weinberg} S. Weinberg, Gravitation and Cosmology, John Wiley \&
Sons, 1972.

\bibitem{salg4} M. Cataldo, J. Crisostomo, S. del Campo, F. Gomez, C.
Quinzacara, P. Salgado, Eur. Phys. J. C 74 (2014) 3087.

\bibitem{garcia} A.A. Garc\'{\i}a, A. Garc\'{\i}a-Quiroz, M. Cataldo, S. del
Campo, Phys. Rev. D 69, 041302(R) (2004).

\bibitem{edelstein} J.D. Edelstein, M. Hassa\"{\i}ne, R. Troncoso, J.
Zanelli, Phys. Lett. B 640 (2006) 278--284.

\bibitem{azcarraga} J. A \ de Azc\'{a}rraga, K. Kamimura, and J. Lukierski,
Phys. Rev. D 83, 124036 (2011).

\bibitem{zan} J. Zanelli, Lecture notes on Chern-Simons (super)gravities,
second edition, arXiv:hep-th/0502193.

\bibitem{irs1} F. Izaurieta, E. Rodriguez, P. Salgado, Lett. Math. Phys. 80
(2007) 127.
\end{thebibliography}
\end{document}